# Unique gap structure and symmetry of the charge density wave in single-layer VSe$_2$

P. Chen,[1,2,3] W.-W. Pai,[4,5] Y.-H. Chan,[6] V. Madhavan,[1,2] M. Y. Chou,[5,6,7] S.-K. Mo,[3] A.-V. Fedorov,[3] and T.-C. Chiang[1,2,5]

[1]Department of Physics, University of Illinois at Urbana-Champaign, 1110 West Green Street, Urbana, Illinois 61801-3080, USA

[2]Frederick Seitz Materials Research Laboratory, University of Illinois at Urbana-Champaign, 104 South Goodwin Avenue, Urbana, Illinois 61801-2902, USA

[3]Advanced Light Source, Lawrence Berkeley National Laboratory, Berkeley, California 94720, USA

[4]Center for Condensed Matter Sciences, National Taiwan University, Taipei 10617, Taiwan

[5]Department of Physics, National Taiwan University, Taipei 10617, Taiwan

[6]Institute of Atomic and Molecular Sciences, Academia Sinica, Taipei 10617, Taiwan

[7]School of Physics, Georgia Institute of Technology, Atlanta, GA 30332, USA

2Abstract

Single layers of transition metal dichalcogenides (TMDCs) are excellent candidates for electronic applications beyond the graphene platform; many of them exhibit novel properties including charge density waves (CDWs) and magnetic ordering. CDWs in these single layers are generally a planar projection of the corresponding bulk CDWs because of the quasi-two-dimensional nature of TMDCs; a different CDW symmetry is unexpected. We report herein the successful creation of pristine single-layer $VSe_2$, which shows a $(\sqrt{7} \times \sqrt{3})$ CDW in contrast to the $(4 \times 4)$ CDW for the layers in bulk $VSe_2$. Angle-resolved photoemission spectroscopy (ARPES) from the single layer shows a sizable $(\sqrt{7} \times \sqrt{3})$ CDW gap of ~100 meV at the zone boundary, a 220 K CDW transition temperature twice the bulk value, and no ferromagnetic exchange splitting as predicated by theory. This robust CDW with an exotic broken symmetry as the ground state is explained via a first-principles analysis. The results illustrate a unique CDW phenomenon in the two-dimensional limit.

Two-dimensional (2D) systems have attracted great interest for their wide varieties of novel properties well suited for advanced electronic applications [1-7]. Specifically, single layers of TMDCs are excellent exploratory candidates that offer huge choices of material types including semiconductors, semimetals, metals, magnetic systems, superconductors, quantum spin Hall insulators, and other topologically nontrivial systems. CDW is a common phenomenon in bulk and ultrathin TMDCs because their (quasi-)2D structures tend to suppress electronic screening while enhance the effects of Fermi surface nesting, electron-phonon coupling, and electron-electron correlation. It is of broad interest as CDWs can entangle or compete with other ordering phenomena. While CDWs have been investigated for decades, there is not yet a single definitive universal theory that conclusively explains all observed phenomena [8-12]. Of all such materials, VSe$_2$ is of special interest because of its unusually long wavelengths for its (4 × 4 × 3) CDW in the bulk below transition temperature $T_C$ = 110 K [13-15]. Prior experiments suggested 3D Fermi surface nesting as a mechanism for the bulk CDW [16-17], but recent theoretical calculations suggested electron-phonon coupling as a key factor [18].

CDWs in single layers of TMDCs are generally a planar projection of the corresponding bulk CDWs because of the quasi-two-dimensional nature of TMDCs; a different CDW symmetry is unexpected [6, 7, 19, 20]. Our study presented herein of single-layer VSe$_2$ shows a ($\sqrt{7} \times \sqrt{3}$) CDW in contrast to the (4 × 4) CDW for the layers in bulk VSe$_2$. This drastically altered symmetry type involving a spontaneous breaking of the triangular crystal symmetry presents an exceptional case. It is also intriguing that the observed 220 K CDW transition temperature in the single layer is twice the bulk value. We looked for, but found no evidence for the theoretically predicted room-temperature ferromagnetic ground state with a large exchange splitting for the single layer [21, 22], which is likely suppressed by the strongly competing CDW order. Our results are contrasted by a recent report of a ferromagnetic response of a single layer capped under Se [23]; the differences suggest strong environmental effects.

In our experiment, VSe$_2$ thin films were grown *in situ* in the integrated MBE/ARPES systems at beamlines 12.0.1 and 10.0.1, Advanced Light Sources, Lawrence Berkeley National Laboratory. Substrates of 6H-SiC(0001) were flash-annealed for multiple cycles to form a well-ordered bilayer graphene on the surface [24]. Films of VSe$_2$ were grown on top of the substrate while the substrate was maintained at 230 °C; the resulting interface was incommensurate but the orientational crystallographic alignment was maintained between the overlayer and the substrate. Bulk VSe$_2$ samples were prepared by cleavage to expose a fresh surface. ARPES measurements were performed with an energy resolution of <20 meV and an angular resolution of 0.2°. The Fermi level is determined by fitting ARPES spectra from a polycrystalline gold sample. For STM/STS measurements, the samples after characterization by ARPES were capped with a 20-nm layer of Se for protection. The protective Se layer was thermally desorbed at 180 °C before measurements using an Omicron LTSTM instrument and freshly flashed tungsten tips. Calculations were performed using the Vienna ab initio package (VASP) [25-27] with the projector augmented wave method [28, 29]. A plane-wave energy cut-off of 320 eV and an 18x18x1 *k*-mesh were employed. The generalized gradient approximation (GGA) with the Perdew-Burke-Ernzerhof (PBE) functional [30] was used for structural optimization of single-layer VSe$_2$. Freestanding films were modeled with a 20-Å vacuum gap between adjacent layers in the supercell. The fully optimized in-plane lattice constant for single layer 1T VSe$_2$ is $a = 3.35$ Å. Phonon calculations were carried out using the supercell method as implemented in the Phonopy package [31]. A higher energy cutoff of 550 eV and a denser 28x28x1 *k*-mesh were used to ensure converged phonon results. Band unfolding was performed using the BandUP code

[32, 33].

The structure of single-layer VSe$_2$ [Fig. 1(a)] in the normal phase consists of a triangular planar net of V atoms sandwiched between two Se atomic layers. Sharp reflection-high-energy-electron diffraction (RHEED) patterns [Fig. 1(b)] reveal a high-quality and well-ordered single layer of VSe$_2$. Bulk VSe$_2$ is formed by vertically stacking the single layers with van der Waals bonding. Topographic images from scanning tunneling microscopy (STM) taken at 77 K reveal a ($\sqrt{7} \times \sqrt{3}$) CDW structure [Fig. 1(c)]. The Fourier transform of an image [Fig. 1(d)] shows intense (1 × 1) primitive lattice spots and ($\sqrt{7} \times \sqrt{3}$) superlattice spots. The 2D Brillouin zones of the (1 × 1) normal phase and the ($\sqrt{7} \times \sqrt{3}$) CDW phase are shown in Fig. 1(e). Spontaneous triangular symmetry breaking in the CDW phase leads to three domain orientations at 120° apart as observed by STM over different parts of the sample. The STM image in Fig. 1(c) was taken from a single domain.

ARPES along the $\overline{\Gamma M}$ direction from a single-layer sample in the CDW phase at 10 K [Fig. 1(f)] shows hole-like bands centered at the $\overline{\Gamma}$ point that are primarily derived from the Se 4$p$ states. The experimental dispersion relations are in excellent agreement with theory [Fig. 1(g)], and the sharpness of the band features indicates excellent sample quality over the macroscopic ARPES probing area (20×100 μm$^2$). While the system is in the CDW state, no folded band features are evident in the data, indicating a weak superlattice distortion in the CDW phase. The single-layer film thickness is determined by the method described in [34] and by STM measurements. Additional ARPES mapping [35] with a variety of photon energies and polarization configurations for sample temperatures as low as 10 K, much lower than the predicted Curie temperature, reveals within our resolution of 20 meV no detectable ferromagnetic exchange band splittings, which are predicted to be on the order of 0.5 eV [21, 22]. We conclude that our pristine films are not ferromagnetic.

Fermi surface maps in $k$ space for the single layer taken at $T = 300$ and 10 K [Figs. 2(a) and 2(b), respectively] reveal differences caused by the CDW transition. The map in the normal phase (300 K) shows a hexagram-shaped hole pocket around the zone center and ellipse-shaped electron pockets centered about the $\overline{M}$ points. With the single layer deep in the CDW state at 10 K, the Fermi surface maps develop dark spots around the elliptical electron pockets, which are indicative of gap formation at the Fermi level. Note that the two long sides of each elliptical pocket in the normal phase are almost straight and nearly parallel. This geometry offers an excellent nesting condition, and the nesting vector is indicated in Fig. 2(a) by a red arrow $\mathbf{q}_1$. Such nesting is expected to give rise to a CDW-modulated lattice with the reciprocal lattice defined by basis vectors $\mathbf{q}_1$ and $\mathbf{q}_2$ as indicated in Fig. 2(a). The length of $\mathbf{q}_1$ closely matches $3/5\overline{\Gamma K}$, which is exactly the condition for the formation of a $(\sqrt{7} \times \sqrt{3})$ CDW-modulated structure; the geometric relationship is illustrated in Fig. 1(e), where $\mathbf{q}_1$ and $\mathbf{q}_2$ are related to the reciprocal lattice vectors $\mathbf{b}_1$ and $\mathbf{b}_2$ for the $(1 \times 1)$ lattice by $\mathbf{b}_1 = 2\mathbf{q}_1 - \mathbf{q}_2$ and $\mathbf{b}_2 = \mathbf{q}_1 + 2\mathbf{q}_2$.

For comparison, measured Fermi surface maps for bulk VSe$_2$ at 300 and 10 K are shown in Figs. 2(c) and 2(d), respectively. The results are overall similar to the single-layer case, but the width of the pocket is smaller by about 16%. This difference upsets the $(\sqrt{7} \times \sqrt{3})$ nesting condition for the single layer, and the CDW modulation of the Fermi surface seen for the single layer is not observed for the bulk sample. Indeed, the CDW structure of bulk VSe$_2$ adopts a $(4 \times 4 \times 3)$ symmetry instead, which has been examined and analyzed in great detail in prior studies [16, 17].

To further characterize the gap for the single layer as revealed by the CDW-induced dark spots in the Fermi contours, we plot measured *E-k* ARPES maps along the $\overline{MK}$ direction for the single layer at 10 and 300 K [Fig. 3(a)]. A V-shaped band is seen, which is primarily derived from the V 3*d* states. The two branches of the V cross the Fermi level in the normal phase, but gaps form in the CDW phase, as illustrated by the ARPES maps obtained by symmetrization in energy about the Fermi level [Fig. 3(b)]. Clearly seen for the CDW phase at 10 K is a gap of ~100 meV at the Fermi level, but not for the normal phase at 300 K. By contrast, the bands around the zone center show no indication of a gap [34]. Calculated band structures for the (1 × 1) normal phase and the energy-optimize ($\sqrt{7} \times \sqrt{3}$) CDW phase are shown in Fig. 3(c), where the ($\sqrt{7} \times \sqrt{3}$) band structure has been "unfolded" back into the (1 × 1) zone for easy comparison. The main difference between the two phases is a CDW gap of 80 meV for the V-shaped V 3*d* band at the Fermi level. This theoretical gap value is close to, but smaller than, the experimental value of 100 meV; however, such density functional calculations do not necessarily yield accurate band gaps. A similar analysis for bulk VSe$_2$ shows no CDW gaps at 10 K for the V-shaped V 3*d* bands along $\overline{MK}$ [Fig. 3(d)] and other high symmetry paths in the Brillouin zone [35]. Our results are consistent with a prior STM study of bulk VSe$_2$ [36]. Although a prior ARPES study of the bulk suggested a "gap" of ~40 meV, the spectral features could be attributed instead to a reduction of the spectral function near the Fermi level [16].

The transition temperature of the CDW is of interest. Figure 4(a) shows the symmetrized energy distribution curves (EDCs) at different temperatures taken at the gap location in momentum space [Fig 3(b)]. The evolution of the line shape from 300 to 10 K is consistent with a peak splitting into two peaks separated by the CDW gap, while simultaneously the peak width diminishes because of reduced thermal broadening at lower temperatures. The energy gap is extracted by fitting each symmetrized EDC with a phenomenological self-energy expression [37]:

$$A(\mathbf{k}, \omega) = \frac{B(\mathbf{k})}{\pi} \frac{\text{Im} \Sigma(\mathbf{k},\omega)}{[\omega - \epsilon(\mathbf{k}) - \text{Re} \Sigma(\mathbf{k},\omega)]^2 + [\text{Im} \Sigma(\mathbf{k},\omega)]^2} \quad (1)$$

$$\Sigma(\mathbf{k}, \omega) = -i\Gamma_1 + \frac{\Delta^2}{[\omega + \epsilon(\mathbf{k}) + i\Gamma_0]} \quad (2)$$

where $A(\mathbf{k}, \omega)$ is the spectra function, $B(\mathbf{k})$ is the corresponding weight, $\Sigma(\mathbf{k}, \omega)$ is the self energy, and $\Delta$ is the gap. An example of the fit to the EDC at 10 K is shown in Fig. 4(b) and the gap is determined to be $101 \pm 5$ meV. The square of the extracted energy gap, plotted as a function of $T$ [Fig. 4(c)], is fitted to a mean-field gap equation for second-order phase transitions [7]:

$$\Delta^2(T) \propto \tanh^2(A\sqrt{\frac{T_C}{T} - 1})\Theta(T_C - T) \quad (3)$$

where $A = 1.19$ is a proportional constant and $\Theta$ is the unit step function. The result of the fit, shown as the blue curve in Fig. 4(c), yields a CDW transition temperature of $T_C = 220 \pm 5$ K. This single-layer transition temperature is twice the bulk transition temperature (110 K). Prior studies of nanoscale and single-layer $VSe_2$ samples have reported conflicting observations of both enhanced and reduced $T_C$ [21, 23, 38], which might be related to different sample preparation methods, varied sample sizes, and different sample environments [36]. Other materials including $TiSe_2$ and $NbSe_2$ also show higher $T_C$ for the single layer than the bulk [5,7], but $VSe_2$ is the only case where the single-layer CDW symmetry is not a simple planer projection of the bulk CDW symmetry.

Based on our first-principles calculations, the total energy for $(\sqrt{7} \times \sqrt{3})$ is lower by 4 meV per chemical unit than the $(4 \times 4)$ structure, after energy minimization for both. The calculated electronic susceptibility and nesting function for the single layer reveal strong features at $\mathbf{q}_1$ and $\mathbf{q}_2$ associated with the $(\sqrt{7} \times \sqrt{3})$ instability [35] in agreement with our ARPES measurements, but these features are generally lacking for the bulk. The computed phonon dispersion relations for the $(1 \times 1)$ phase show imaginary modes at the $(\sqrt{7} \times \sqrt{3})$ CDW $\mathbf{q_1}$ wavevectors, but they also show imaginary modes at wavevectors between $\bar{\Gamma}$ and $\bar{M}$ related to a $(4 \times 4)$ distortion [35]. Evidently, the imaginary modes at $\mathbf{q_1}$ dominate, causing the formation of the $(\sqrt{7} \times \sqrt{3})$ structure as the ground state.

The total evidence based on both experiment and theory indicates that single-layer VSe$_2$ is a Peierls system for which nearly perfect Fermi surface nesting over long segments of the Fermi contour leads to a robust CDW. The resulting $(\sqrt{7} \times \sqrt{3})$ symmetry breaks the three-fold symmetry of the $(1 \times 1)$ normal phase, and the CDW is characterized by strong features in the electronic susceptibility and nesting function and a corresponding phonon instability. Nearly ideal nesting is extremely rare in 2D and 3D systems. CDWs in single layers of TMDCs are generally a planar projection of the corresponding bulk CDWs. Single-layer VSe$_2$ is a unique case; its $(\sqrt{7} \times \sqrt{3})$ CDW in contrast to the $(4 \times 4)$ CDW for the layers in bulk VSe$_2$ involves a spontaneous breaking of the triangular crystal symmetry. This different behavior can be attributed to small differences in the Fermi contours in going from the single layer to the bulk. The sensitivity illustrates that closeness of Fermi surface nesting in 2D may have a strong influence on the energetics of the system, and it explains why the CDW transition temperature for the single layer is substantially higher than the bulk CDW. An interesting prospect for future research is to modify the Fermi contours in single-layer VSe$_2$ by, for example, alloying, strain, and interfacial engineering, and thereby to create a 2D ferromagnet [39, 40] or other exotic phases. Our experiments also provide impetus for a more comprehensive theoretical analysis of the competition between CDW and magnetic orders [41].

**Acknowledgments**

This work is supported by the U.S. Department of Energy, Office of Science, Office of Basic Energy Sciences, Division of Materials Science and Engineering, under Grant No. DE-FG02-07ER46383 (TCC), the National Science Foundation under Grant No. EFMA-1542747 (MYC), the Ministry of Science and Technology of Taiwan under Grant No. 103-2923-M-002-003-MY3 (WWP), and the Gordon and Betty Moore Foundation's EPiQS Initiative through Grant GBMF4860 (VM). The Advanced Light Source is supported by the Director, Office of Science, Office of Basic Energy Sciences, of the U.S. Department of Energy under Contract No. DE-AC02-05CH11231. The work at Academia Sinica is supported by a Thematic Project. We thank Alpha T. N'Diaye for insightful discussions.

Note added.— During the review of this paper, a related study by Feng *et al*., with a later submission date, appeared in print [42], which reported a CDW gap in single layer $VSe_2$ and an enhanced CDW transition temperature. Their findings are consistent with ours qualitatively. However, we find a different value of the transition temperature and only partial gaps.

**References**

[1] A. H. Castro Neto, F. Guinea, N. M. R. Peres, K. S. Novoselov, and A. K. Geim, Rev. Mod. Phys. **81**, 109 (2009).

[2] Q. H. Wang, K. Kalantar-Zadeh, A. Kis, J. N. Coleman, and M. S. Strano, Nat. Nanotech. **7**, 699 (2012).

[3] Y. Zhang, K. He, C.-Z. Chang, C.-L. Song, L.-L. Wang, X. Chen, J.-F. Jia, Z. Fang, X. Dai, W.-Y. Shan *et al.*, Nat. Phys. **6**, 712 (2010).

[4] Q.-Y. Wang, Z. Li, W.-H. Zhang, Z.-C. Zhang, J.-S. Zhang, W. Li, H. Ding, Y.-B. Ou, P. Deng, K. Chang *et al.*, Chin. Phys. Lett. **29**, 037402 (2012).

[5] X. Xi, L. Zhao, Z. Wang, H. Berger, L. Forró, J. Shan, and K. F. Mak, Nat. Nanotech. **10**, 765 (2015).

[6] M. M. Ugeda, A. J. Bradley, Y. Zhang, S. Onishi, Y. Chen, W. Ruan, C. Ojeda-Aristizabal, H. Ryu, M. T. Edmonds, H.-Z. Tsai *et al.*, Nature Phys. **12**, 92 (2016).

[7] P. Chen, Y.-H. Chan, X.-Y. Fang, Y. Zhang, M.Y. Chou, S.-K. Mo, Z. Hussain, A.-V. Fedorov, and T.-C. Chiang, Nat. Commun. **6**, 8943 (2015).

[8] K. Rossnagel, J. Phys. Condens. Matter **23**, 213001 (2011).

[9] S. V. Borisenko, A. A. Kordyuk, A. N. Yaresko, V. B. Zabolotnyy, D. S. Inosov, R. Schuster, B. Büchner, R. Weber, R. Follath, L. Patthey, and H. Berger, *Phys Rev. Lett.* **100**, 196402 (2008).

[10] M. Holt, P. Zschack, H. Hong, M. Y. Chou, and T.-C. Chiang, Phys. Rev. Lett. **86**, 3799 (2001).

[11] Th. Straub, Th. Finteis, R. Claessen, P. Steiner, S. Hüfner, P. Blaha, C. S. Oglesby, and E. Bucher, Phys. Rev. Lett. **82**, 4504 (1999).

[12] T. Valla, A. V. Fedorov, P. D. Johnson, P-A. Glans, C. McGuinness, K. E. Smith, E. Y. Andrei, and H. Berger, Phys. Rev. Lett. **92**, 086401 (2004).

[13] K. Tsutsumi, Phys. Rev. B **26**, 5756 (1982).

[14] A. H. Thompson, and B. G. Silbernagel, Phys. Rev. B **19**, 3420 (1979).

[15] D. J. Eaglesham, R. L. Withers, and D. M. Bird, J. Phys. C **19**, 359 (1986).

[16] K. Terashima, T. Sato, H. Komatsu, T. Takahashi, N. Maeda, and K. Hayashi, Phys. Rev. B **68**, 155108 (2003).

[17] V. N. Strocov, M. Shi, M. Kobayashi, C. Monney, X. Wang, J. Krempasky, T. Schmitt, L. Patthey, H. Berger, and P. Blaha, Phys. Rev. Lett. **109**, 086401 (2012).

[18] D. Zhang, J. Ha, H. Baek, Y.-H. Chan, F. D. Natterer, A. F. Myers, J. D. Schumacher, W. G. Cullen, A. V. Davydov, Y. Kuk *et. al.*, Phys. Rev. Materials **1**, 024005 (2017).

[19] H. Ryu, Y. Chen, H. Kim, H.-Z. Tsai, S. Tang, J. Jiang, F. Liou, S. Kahn, C. Jia, A. A. Omrani *et al.*, Nano Lett. **18**, 689 (2018).

[20] M. Calandra, I. I. Mazin, and F. Mauri, Phys. Rev. B **80**, 241108 (2009).

[21] K. Xu, P. Chen, X. Li, C. Wu, Y. Guo, J. Zhao, X. Wu, and Y. Xie, Angew. Chem., Int. Ed. **52**, 10477 (2013).

[22] Y. Ma, Y. Dai, M. Guo, C. Niu, Y. Zhu, and B. Huang, ACS nano **6**, 1695 (2012).

[23] M. Bonilla, S. Kolekar, Y. Ma, H. Coy Diaz, V. Kalappattil, R. Das, T. Eggers, H. R. Gutierrez, M.-H. Phan, and M. Batzill, Nat. Nanotech. doi:10.1038/s41565-018-0063-9.

[24] Q. Y. Wang, W. H. Zhang, L. L. Wang, K. He, X. C. Ma, and Q. K. Xue, J. Phys.: Condens. Matter **25**, 095002 (2013).

[25] G. Kresse, and J. Hafner, Phys. Rev. B **48**, 13115 (1993).

[26] G. Kresse, and J. Furthmüller, Comput. Mater. Sci. **6**, 15(1996).

[27] G. Kresse, and J. Furthmüller, Phys. Rev. B **54**, 11169 (1996).

[28] P. E. Blöchl, Phys. Rev. B **50**, 17953 (1994).

[29] G. Kresse, and D. Joubert, Phys. Rev. B **59**, 1758 (1999).

[30] J. P. Perdew, K. Burke, and M. Ernzerhof, Phys. Rev. Lett. **77**, 3865 (1996).

[31] A. Togo, and I. Tanaka, Scr. Mater. **108**, 1 (2015).

[32] P. V. C. Medeiros, S. Stafström, and J. Björk, Phys. Rev. B **89**, 041407(R) (2014).

[33] P. V. C. Medeiros, S. S. Tsirkin, S. Stafström, and J. Björk, Phys. Rev. B **91**, 041116(R) (2015).

[34] P. Chen, Y.-H. Chan, M.-H. Wong, X.-Y. Fang, M. Y. Chou, S.-K. Mo, Z. Hussain, A.-V. Fedorov, and T.-C. Chiang, Nano Lett. **16**, 6331 (2016).

[35] See Supplemental Material at [URL will be inserted by publisher] for additional data and details. It cites Refs. [16-18, 20, 34, 43].

[36] Á. Pásztor, A. Scarfato, C. Barreteau, E. Giannini, and C. Renner, 2D Mater. **4,** 041005 (2017).

[37] M. R. Norman, M. Randeria, H. Ding, and J. C. Campuzano, Phys. Rev. B **57**, R11093 (1998).

[38] J. Yang, W. Wang, Y. Liu, H. Du, W. Ning, G. Zheng, C. Jin, Y. Han, N. Wang, Z. Yang, M. Tian, and Y. Zhang. Appl. Phys. Lett. **105** 063109 (2014).

[39] B. Huang, G. Clark, E. Navarro-Moratalla, D. R. Klein, R. Cheng, K. L. Seyler, D. Zhong, E. Schmidgall, M. A. McGuire, D. H. Cobden *et. al.*, Nature **546**, 270 (2017).

[40] C. Gong, L. Li, Z. Li, H. Ji, A. Stern, Y. Xia, T. Cao, W. Bao, C. Wang, Y. Wang *et. al.*, Nature **546**, 265 (2017).

[41] A. O. Fumega, and V. Pardo, arXiv:1804.07102.

[42] J. Feng, D. Biswas, A. Rajan, M. D. Watson, F. Mazzola, O. J. Clark, K. Underwood, I. Marković, M. McLaren, A. Hunter *et. al.*, Nano Lett. **18**, 4493 (2018).

[43] M. D. Johannes, I. I. Mazin, and C. A. Howells, Phys. Rev. B **73**, 205102 (2006).

FIG 1. Film structure, electronic band structure, and STM images of single-layer VSe$_2$. (a) Atomic structure of single-layer 1T VSe$_2$. (b) A RHEED pattern taken at room temperature after single-layer VSe$_2$ film growth. (c) A STM topographic image taken from single-layer VSe$_2$ at 77 K revealing ($\sqrt{7} \times \sqrt{3}$) CDW modulation, as indicated by the blue solid line unit cell. The experimental conditions are: size 9.9 x 6.2 nm, sample bias -0.5 V, and tunneling current 0.5 nA. (d) Fourier transform of the STM image in (c). The (1 × 1) zone and the ($\sqrt{7} \times \sqrt{3}$) reciprocal lattice vectors are indicated. (e) Brillouin zones of the (1 × 1) and ($\sqrt{7} \times \sqrt{3}$) structures are outlined in blue and red, respectively. **q**$_1$ and **q**$_2$ are ($\sqrt{7} \times \sqrt{3}$) primitive reciprocal lattice vectors. (f) An ARPES map along $\overline{\Gamma M}$ taken at 10 K. (g) Calculated band dispersion relations for the (1 × 1) structure without spin polarization.

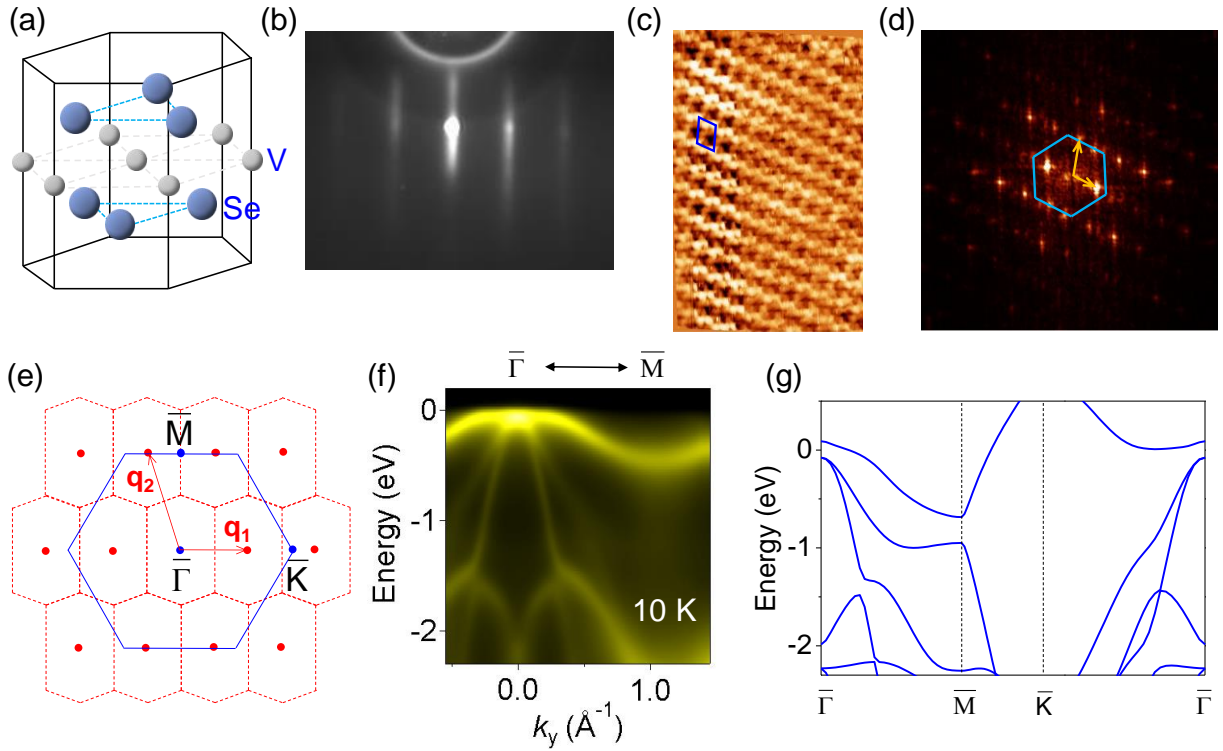

FIG 2. Fermi surfaces of single-layer and bulk VSe$_2$ in the CDW and normal phases. (a) Measured Fermi surface contours of single-layer VSe$_2$ in the normal phase at 300 K obtained by integrating the ARPES intensity over ±10 meV around the Fermi level. The blue hexagon indicates the first Brillouin zone; $\mathbf{q}_1$ and $\mathbf{q}_2$ indicate the ($\sqrt{7} \times \sqrt{3}$) CDW wave vectors. (b) Fermi contours for the CDW phase at 10 K. Dark spots spanned by $\mathbf{q}_1$ are caused by nesting and CDW gap formation. [(c) and (d)] Fermi surface maps of bulk VSe$_2$ in the normal phase at 300 K and the CDW phase at 10 K, respectively.

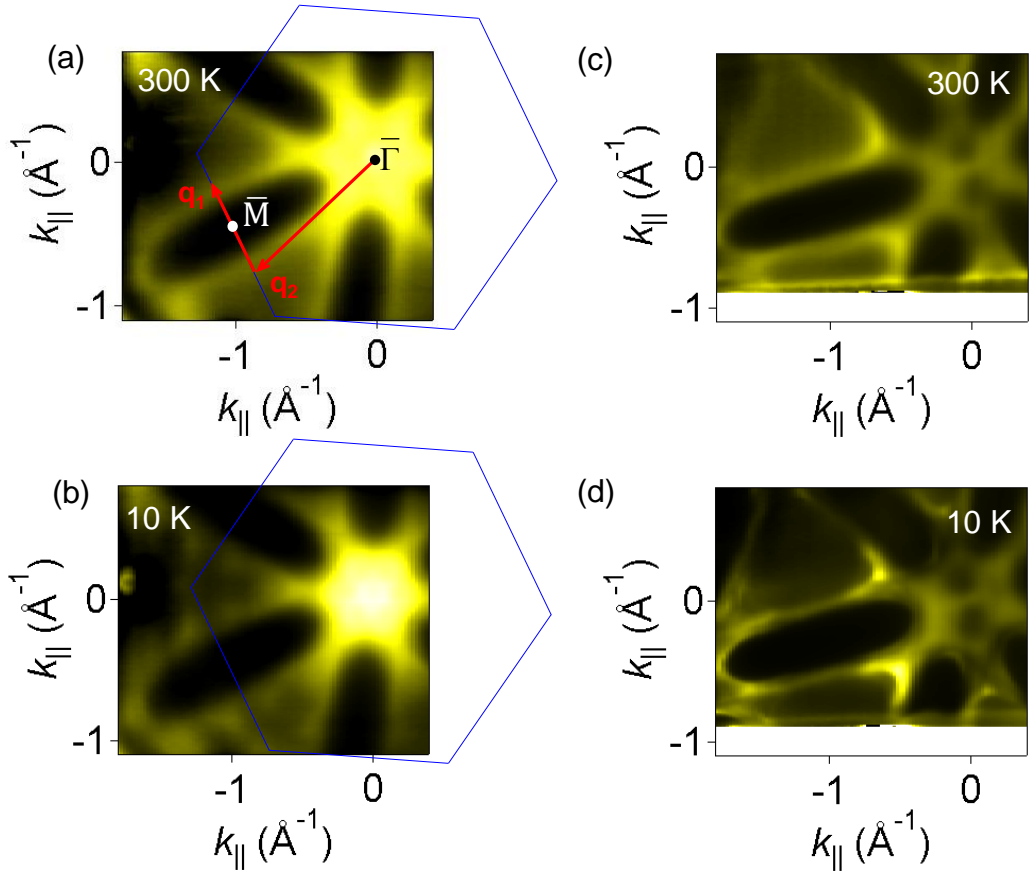

FIG 3. CDW gap opening in single-layer VSe$_2$. (a) ARPES spectra along $\overline{MK}$ for single-layer VSe$_2$ in the normal phase at 300 K (upper panel) and the CDW phase at 10 K (lower panel). (b) Corresponding ARPES maps symmetrized in energy about the Fermi level show a gap in the CDW phase. The nesting vector q$_1$ is shown as a red arrow. (c) Calculated band structure in the normal and CDW phases. The CDW gap is indicated. A zoom-in plot is shown at the bottom to show details. (d) ARPES spectra along the MK direction for bulk VSe$_2$ at 10K (top panel). The symmetrized map (bottom panel) shows no gap.

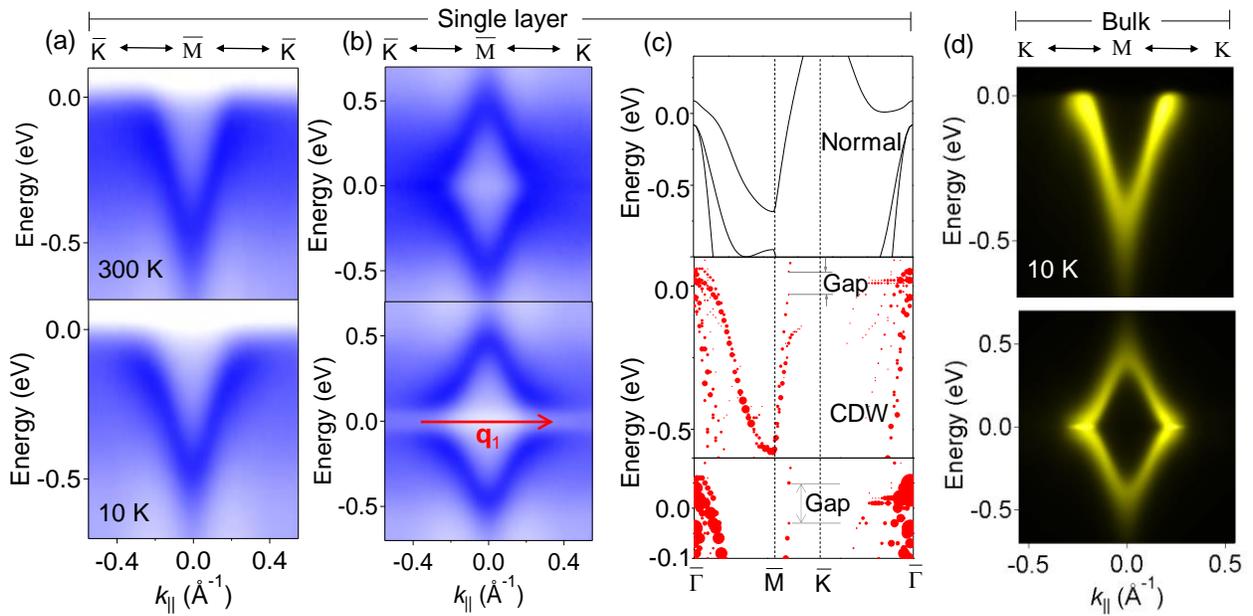

FIG 4. Temperature dependence of the CDW gap and the transition temperature. (a) Symmetrized EDCs at the gap location at selected temperatures between 10 and 300 K. The evolution of the line shape is consistent with a peak splitting into two peaks separated by the CDW gap, while simultaneously the peak width diminishes because of reduced broadening at lower temperatures. The energy gap is extracted by fitting each symmetrized EDC with a phenomenological self-energy formula [26]. (b) An example of the fit shown as a red curve for the EDC obtained at 10 K. (c) The extracted temperature dependence of the square of the CDW gap. The blue curve is a fit using a mean-field equation. Transition temperature $T_c$ is labeled.

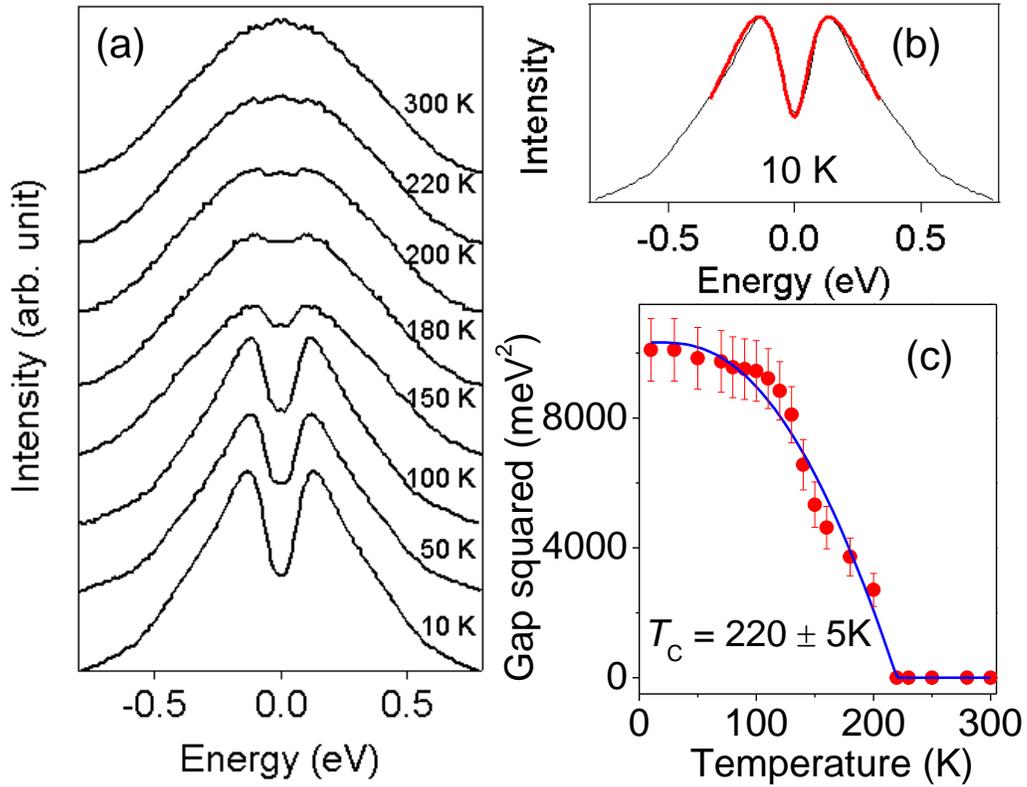